\documentclass[a4paper,showpacs,showkeys,nofootinbib,aps]{revtex4}
\usepackage{amsmath}
\usepackage{epsfig}

\begin{document}
\title{Single Spin Asymmetry\\
in Open Charm Photoproduction and Decay\\
as a Test of pQCD\\}
\author{N.Ya.~Ivanov}
\email{nikiv@yerphi.am}
\affiliation{Yerevan Physics Institute, Alikhanian Br.2, 0036 Yerevan, Armenia}
\author{P.E.~Bosted}
\email{bosted@slac.stanford.edu}
\affiliation{University of Massachusetts, Amherst,  Massachusetts 01003}
\author{K.~Griffioen}
\email{griff@physics.wm.edu}
\affiliation{College of William and Mary, Williamsburg, Virginia 23187}
\author{S.E.~Rock}
\email{ser@slac.stanford.edu}
\affiliation{University of Massachusetts, Amherst,  Massachusetts 01003}
\begin{abstract}
\noindent We analyze the possibility of measuring the single spin
asymmetry (SSA) in open charm production by linearly polarized
photons in the planned E160/E161 experiments at SLAC using the
inclusive spectra of secondary (decay) leptons. In leading order 
pQCD, the SSA in the azimuthal distribution of the charged
decay lepton is predicted to be about $0.2$ for SLAC kinematics.
Our calculations show that the SSA in the decay lepton
distribution is well defined in pQCD: it is stable both
perturbatively and parametrically, and practically insensitive to
theoretical uncertainties in the charm semileptonic decays.
Nonperturbative contributions to the leptonic SSA due to the gluon
transverse motion in the target, heavy quark fragmentation and 
Fermi motion of the $c$-quark inside the $D$-meson are
predicted to be about $10\%$. We conclude that measurements of the
azimuthal asymmetry in the secondary lepton distribution would
provide a good test of the applicability of pQCD to open charm production
at energies of fixed target experiments. Our analysis of the SLAC
experimental conditions shows that this SSA can be measured in
E160/E161 with an accuracy of about ten percent.\\
\end{abstract}
\maketitle
\section{ Introduction}
In the framework of perturbative QCD, the basic spin-averaged
characteristics of heavy flavor hadro-, photo- and
electroproduction are known exactly up to the next-to-leading
order (NLO). During the last ten years, these NLO results have
been widely used for a phenomenological description of available
data (for a review see \cite{1}). At the same time, the key
question remains open: How to test the applicability of QCD
at fixed order to heavy quark production? The problem is
twofold. On the one hand, the NLO corrections are large; they
increase the leading order (LO) predictions for both charm and
bottom production cross sections by approximately a factor of two.
For this reason, one could expect that higher-order corrections,
as well as nonperturbative contributions, can be essential, 
especially for the $c$-quark case. On the other hand, it is very difficult 
to compare pQCD predictions for spin-averaged cross sections
with experimental data directly, without additional assumptions,  
because of a high sensivity of the
theoretical calculations to standard uncertainties in the input
QCD parameters. The total uncertainties associated with the
unknown values of the heavy quark mass, $m_{Q}$, the factorization
and renormalization scales, $\mu _{F}$ and  $\mu _{R}$, $\Lambda
_{QCD}$ and the parton distribution functions are so large that
one can only estimate the order of magnitude of the pQCD
predictions for total cross sections at fixed target energies
\cite{2,3}.

In recent years, the role of higher-order corrections has been
extensively investigated in the framework of the soft gluon
resummation formalism. For a review see Ref.\cite{4}. Formally
resummed cross sections are ill-defined due to the Landau pole
contribution, and a few prescriptions have been proposed  to avoid
the renormalon ambiguities \cite{5,6,7}. Unfortunately, numerical
predictions for the heavy quark production cross sections can
depend significantly on the choice of resummation prescription
\cite{8}. Another open question, also closely related to convergence 
of the perturbative series, is the role of subleading 
contributions which are not, in principle, under control of the
resummation procedure \cite{8,8L}.

For this reason, it is of special interest to study those
observables that are well-defined in pQCD. An nontrivial example
of such an observable is proposed in  \cite{9,10}, where the charm
and bottom production by linearly polarized photons,
\begin{equation}
\gamma ^{\uparrow }+N\rightarrow Q+X[\overline{Q}],  \label{1}
\end{equation}
was considered\footnote{The well-known examples are the shapes of
differential cross sections of heavy flavor production which are
sufficiently stable under radiative corrections.}. In particular,
the single spin asymmetry (SSA) parameter, $A_{Q}(p_{QT})$, which
measures the parallel-perpendicular asymmetry in the quark
azimuthal distribution,
\begin{equation}
\frac{{\text{d}}^{2}\sigma _{Q}}{\text{d}p_{QT}\text{d}\varphi _{Q}}
(p_{QT},\varphi _{Q})=\frac{1}{2\pi }\frac{\text{d}\sigma _{Q}^{\rm{unp}}}
{\text{d}p_{QT}}(p_{QT})\left[ 1+A_{Q}(p_{QT}){\cal P}_{\gamma }\cos
2\varphi _{Q}\right],   \label{2}
\end{equation}
where
\begin{equation}
A_{Q}(p_{QT})=\frac{1}{{\cal P}_{\gamma }}\frac{\text{d}^{2}\sigma
_{Q}(p_{QT},\varphi _{Q}=0)-\text{d}^{2}\sigma _{Q}(p_{QT},\varphi _{Q}=\pi
/2)}{\text{d}^{2}\sigma _{Q}(p_{QT},\varphi _{Q}=0)+\text{d}^{2}\sigma
_{Q}(p_{QT},\varphi _{Q}=\pi /2)},  \label{3}
\end{equation}
has been calculated. In (\ref{2}) and (\ref{3}), $\frac{\text{d}\sigma _{Q}^{
\rm{unp}}}{\text{d}p_{QT}}$ is the unpolarized cross section, d$%
^{2}\sigma _{Q}(p_{QT},\varphi _{Q})= \frac{\text{d}^{2}\sigma _{Q}}{%
\text{d}p_{QT}\text{d}\varphi _{Q}}(p_{QT},\varphi _{Q})$, ${\cal P}%
_{\gamma }$ is the degree of linear polarization of the incident
photon beam and $\varphi _{Q}$ is the angle between the beam
polarization direction and the observed quark transverse momentum,
$p_{QT}$. The following remarkable properties of the SSA,
$A_{Q}(p_{QT})$, have been observed  \cite{9}:

\begin{itemize}
\item  The azimuthal asymmetry (\ref{3}) is of leading twist; in a
wide kinematical region, it is predicted to be about $0.2$ for
both charm and bottom quark production.
\item  At energies sufficiently above the production threshold, the
LO predictions for $A_{Q}(p_{QT})$ are insensitive (to within few
percent) to uncertainties in the QCD input parameters.
\item   Nonperturbative corrections to the $b$-quark azimuthal asymmetry
are negligible. Because of the smallness of the $c$-quark mass,
the analogous corrections to $A_{c}(p_{QT})$ are larger; they are
of the order of $20\%$.
\end{itemize}

In Ref. \cite{10}, radiative corrections to the integrated cross
section,
\begin{equation}
\frac{\text{d}\sigma_{Q}}{\text{d}\varphi_{Q}} (S,\varphi_{Q} )=
\frac{\sigma_{Q}^{\text{{\rm unp}}}(S)}{2\pi }[1+A_{Q}(S){\cal
P}_{\gamma } \cos 2\varphi_{Q}],   \label{A1}
\end{equation}
have been investigated in the soft-gluon approximation. (In Eq. 
(\ref{A1}), $\sqrt{S}$ is the center-of-mass energy of the process
(\ref{1})). Calculations \cite{10} indicate a high perturbative
stability of the pQCD predictions for $A_{Q}(S)$. In particular,

\begin{itemize}
\item  At the next-to-leading logarithmic (NLL) level, the NLO and NNLO
predictions for $A_{Q}(S)$ affect the LO results by less than
$1\%$ and $2\%$, respectively.
\item  Computations of the higher order contributions (up to the 6th order
in $\alpha _{s}$) to the NLL accuracy lead only to a few percent
corrections to the Born result for $A_{Q}(S)$. This implies that
large soft-gluon contributions to the spin-dependent and
unpolarized cross sections cancel each other in Eq. (\ref{3}) with a
good accuracy.
\end{itemize}

Therefore, contrary to the production cross sections, the single spin
asymmetry in heavy flavor photoproduction is an observable
quantitatively well defined in pQCD: it is stable, both
parametrically and perturbatively, and insensitive to
nonperturbative corrections. Measurements of the SSA in bottom 
photoproduction would provide an ideal test of pQCD. Data on the 
D-meson azimuthal distribution would make it possible to clarify the 
role of subleading twist contributions \cite{9,B2}.

Concerning the experimental aspects, the azimuthal asymmetry in
charm photoproduction can be measured at SLAC where a coherent
bremsstrahlung beam of linearly polarized photons with energies up
to 40 GeV will be available soon \cite{11}. In the approved  experiments E160
and E161, charm production will be investigated
using the spectra of decay muons:
\begin{equation}
\gamma ^{\uparrow }+N\rightarrow c+X[\overline{c}]\rightarrow \mu ^{+}+X.
\label{4}
\end{equation}

In this paper, we analyze the possibility to measure the SSA in heavy quark
photoproduction using the decay lepton spectra. We calculate the SSA in the
decay lepton azimuthal distribution:
\begin{equation}
\frac{\text{d}^{2}\sigma _{\ell }}{\text{d}p_{\ell T}\text{d}\varphi _{\ell }%
}(p_{\ell T},\varphi _{\ell })=\frac{1}{2\pi }\frac{\text{d}\sigma _{\ell }^{%
{\rm{unp}}}}{\text{d}p_{\ell T}}(p_{\ell T})\left[ 1+A_{\ell
}(p_{\ell T}){\cal P}_{\gamma }\cos 2\varphi _{\ell }\right] ,  \label{5}
\end{equation}
where $\varphi _{\ell }$ is the angle between the photon polarization
direction and the decay lepton transverse momentum, $p_{\ell T}$. Our main
results can be formulated as follows:

\begin{itemize}
\item  The SSA transferred from the decaying $c$-quark to the decay muon is
large in the SLAC kinematics; the ratio
$A_{\ell}(p_{T})/A_{c}(p_{T})$ is about $90\%$ for $p_{T}>$1 GeV.
\item  pQCD predictions for $A_{\ell }(p_{\ell T})$ are also stable, both
perturbatively and parametrically.
\item  Nonperturbative corrections to $A_{\ell }(p_{\ell T})$ due to the gluon
transverse motion in the target and the $c$-quark fragmentation
are small; they are about $10\%$ for $p_{\ell T}>$1 GeV.
\item  The SSA in Eq. (\ref{5}) depends weakly on theoretical uncertainties in
the charm semileptonic decays\footnote{For a review see Ref.
\cite{21}.}, $c\rightarrow \ell ^{+}\nu _{\ell }X_{q}$ $(q=d,s)$.
In particular,
\begin{itemize}
\item Contrary to the the production cross sections, the asymmetry
$A_{\ell }(p_{\ell T})$ is practically insensitive to the
unobserved stange quark mass, $m_s$, for $p_{\ell T}>$1 GeV.
\item The bound state effects due to the Fermi motion of the $c$-quark inside
the $D$-meson have only a small impact on $A_{\ell }(p_{\ell T})$,
in practically the whole region of $p_{\ell T}$.
\end{itemize}
\end{itemize}

We conclude that the SSA in the decay lepton azimuthal distribution (\ref
{5}) is also well-defined in the framework of perturbation theory and can be
used as a good test of pQCD applicability to open charm production.

The paper is organized as follows. In Section 2 we analyze the
properties of the decay lepton azimuthal distribution at leading
order. We also give the physical explanation of the fact that 
pQCD predictions for the SSA are approximately the same at LO and
at NLO. The details of our calculations of radiative corrections
are too long to be presented in this paper; they will be reported
separately in a forthcoming publication \cite{14}. In Section 3 we
discuss the nonperturbative contributions to $A_{\ell }(p_{\ell
T})$ caused by the Peterson fragmentation, the $k_{T}$-kick and
the Fermi motion of the heavy quark inside the heavy hadron.
Section 4 contains experimental considerations. We discuss the 
conditions needed to measure the SSA at SLAC, estimate background
contributions and expected errors.

\section{pQCD Predictions for SSA}

\subsection{Partonic Cross Sections}

At the Born level, the only partonic subprocess which is
responsible for the reaction (\ref{4}) is the heavy quark
production by photon-gluon fusion,
\begin{equation}
\gamma ^{\uparrow }(k_{\gamma })+g(k_{g})\rightarrow Q(p_{Q})+\overline{Q}%
(p_{\overline{Q}})\rightarrow \ell (p_{\ell })+\nu _{\ell }+q+\overline{Q},
\label{6}
\end{equation}
with subsequent decay $c\rightarrow \ell ^{+}\nu _{\ell }q$
$(q=d,s)$ in the charm case and $b\rightarrow \ell
^{-}\overline{\nu }_{\ell }q$ $(q=u,c)$ in the bottom one. To
calculate distributions of final particles appearing in a process
of production and subsequent decay, it is useful to adopt the
narrow-width approximation,
\begin{equation}
\frac{1}{\left( p_{Q}^{2}-m_{Q}^{2}\right) ^{2}+\Gamma _{Q}^{2}m_{Q}^{2}}%
\rightarrow \frac{\pi }{\Gamma _{Q}m_{Q}}\delta \left(
p_{Q}^{2}-m_{Q}^{2}\right),  \label{7}
\end{equation}
with $\Gamma _{Q}$ the total width of the heavy quark. Corrections
to this approximation are negligibly small in both charm and
bottom cases since they have a relative size ${\cal O}(\Gamma
_{Q}/m_{Q})$.

In the case of the linearly polarized photon, the heavy quark
produced in the reaction (\ref{6}) is unpolarized. For this
reason, the single-inclusive cross section for the decay lepton
production in (\ref{6}) is a simple convolution:
\begin{equation}
E_{\ell }\frac{\text{d}^{3}\hat{\sigma}_{\ell }}{\text{d}^{3}p_{\ell }}({%
\vec{p}}_{\ell })=\frac{1}{\Gamma _{Q}}\int \frac{\text{d}^{3}p_{Q}}{E_{Q}}%
\frac{E_{Q}\text{d}^{3}\hat{\sigma}_{Q}}{\text{d}^{3}p_{Q}}({\vec{p}}_{Q})%
\frac{E_{\ell }\text{d}^{3}\Gamma _{\rm{sl}}}{\text{d}^{3}p_{\ell }}%
(p_{\ell }\cdot p_{Q}).  \label{8}
\end{equation}

At leading order, ${\cal O}(\alpha_{em} \alpha_{s})$, the
$\varphi_{Q}$-dependent cross section for heavy flavor production,
\begin{eqnarray}
\frac{E_{Q}\text{d}^{3}\hat{\sigma}_{Q}}{\text{d}^{3}p_{Q}}({\vec{p}}_{Q})
&\equiv &\frac{2s\text{d}^{3}\hat{\sigma}_{Q}}{\text{d}u_{1}\text{d}t_{1}%
\text{d}\varphi _{Q}}\left( s,t_{1},u_{1},\varphi _{Q}\right)  \nonumber \\
&=&\frac{1}{\pi s}\left[ B_{Q}\left( s,t_{1},u_{1}\right) +\Delta
B_{Q}\left( s,t_{1},u_{1}\right) {\cal P}_{\gamma }\cos 2\varphi
_{Q}\right],  \label{9}
\end{eqnarray}
is given by \cite{9,A0}
\begin{equation}
B_{Q}\left( s,t_{1},u_{1}\right) =\pi e_{Q}^{2}\alpha _{em}\alpha _{s}\left[
\frac{t_{1}}{u_{1}}+\frac{u_{1}}{t_{1}}+\frac{4m_{Q}^{2}s}{t_{1}u_{1}}\left(
1-\frac{m_{Q}^{2}s}{t_{1}u_{1}}\right) \right] \delta \left(
s+t_{1}+u_{1}\right) ,  \label{10}
\end{equation}
\begin{equation}
\Delta B_{Q}\left( s,t_{1},u_{1}\right) =\pi e_{Q}^{2}\alpha _{em}\alpha
_{s}\left[ \frac{4m_{Q}^{2}s}{t_{1}u_{1}}\left( 1-\frac{m_{Q}^{2}s}{%
t_{1}u_{1}}\right) \right] \delta \left( s+t_{1}+u_{1}\right) ,
\label{11}
\end{equation}
with $e_Q$ the quark charge in units of electromagnetic coupling
constant. The corresponding partonic kinematical variables are
\begin{eqnarray}
u_{1} &=&\left( k_{\gamma }-p_{Q}\right) ^{2}-m_{Q}^{2}=-E_{Q}\sqrt{s}%
(1-\alpha \cos \theta _{Q}),\qquad s=\left( k_{\gamma }+k_{g}\right) ^{2},
\nonumber \\
t_{1} &=&\left( k_{g}-p_{Q}\right) ^{2}-m_{Q}^{2}=-E_{Q}\sqrt{s}(1+\alpha
\cos \theta _{Q}),\qquad \alpha =\sqrt{1-m_{Q}^{2}/E_{Q}^{2}},  \label{12}
\end{eqnarray}
with $\theta _{Q}$ and $E_{Q}$ the heavy quark polar angle and energy in the
$\gamma g$ center-of-mass system.

At the tree level, the invariant width of the semileptonic decay $%
Q\rightarrow \ell \nu _{\ell }X_{q}$ can be written as
\begin{equation}
\frac{E_{\ell }\text{d}^{3}\Gamma _{\rm{sl}}}{\text{d}^{3}p_{\ell }}%
(x)\equiv I_{\rm{sl}}(x)=\frac{G_{F}^{2}m_{Q}^{3}}{(2\pi )^{4}}\frac{x%
\left( 1-x-\delta ^{2}\right) ^{2}}{1-x}\times \left\{{\left|
V_{CKM}\right|} ^{2}, \quad Q=c\atop\frac{\left| V_{CKM}\right| ^{2}}{6(1-x)}%
(3-2x+\delta ^{2}\frac{3-x}{1-x}),  \quad Q=b\right. \label{13}
\end{equation}
Here $V_{CKM}$ denotes the corresponding element of the
Cabbibo-Kobayashi--Maskawa matrix, $G_{F}$ is the Fermi constant, $\delta
=m_{q}/m_{Q}$ and
\begin{equation}
x=\frac{2(p_{\ell }\cdot p_{Q})}{m_{Q}^{2}}=\frac{2E_{\ell }E_{Q}}{m_{Q}^{2}}%
(1-\alpha \cos \theta _{\ell Q}),  \label{14}
\end{equation}
where $E_{\ell }$ is the lepton energy and $\theta _{\ell Q}$ is the angle
between the lepton and heavy quark momenta in the $\gamma g$ center-of-mass
system.

To perform the angular integrations in Eq. (\ref{8}), we use the cosine theorem:
\begin{equation}
\cos \theta _{Q}=\cos \theta _{\ell }\cos \theta _{\ell Q}-\sin \theta
_{\ell }\sin \theta _{\ell Q}\cos \varphi _{\ell Q},  \label{15}
\end{equation}
where $\varphi _{\ell Q}$ is the azimuth of the lepton momentum in the frame
where the third axis is directed along the heavy quark momentum.
Substituting d$^{2}\Omega _{Q}\rightarrow $d$^{2}\Omega _{\ell Q}=$d$\cos
\theta _{\ell Q}$d$\varphi _{\ell Q}$ in Eq. (\ref{8}), we obtain the following
master formula for the decay lepton azimuthal distribution:
\begin{eqnarray}
E_{\ell }\frac{\text{d}^{3}\hat{\sigma}_{\ell }}{\text{d}^{3}p_{\ell }}({%
\vec{p}}_{\ell }) &=&\frac{1}{\pi s\Gamma _{Q}}\int\limits_{{\tilde E}_{Q}}^{%
\sqrt{s}/2}\left| {\vec{p}}_{Q}\right| \text{d}E_{Q}\int%
\limits_{\cos{\tilde\theta} _{\ell Q}}^{1}\text{d}\cos \theta _{\ell Q}I_{\rm{sl}%
}(x)\int\limits_{0}^{2\pi }\text{d}\varphi _{\ell Q}  \label{16} \\
&&\times \left[ B_{Q}\left( t_{1},u_{1}\right) +\left( 1-\frac{2\sin
^{2}\theta _{\ell Q}\sin ^{2}\varphi _{\ell Q}}{\sin ^{2}\theta _{Q}}\right)
\Delta B_{Q}\left( t_{1},u_{1}\right) {\cal P}_{\gamma }\cos 2\varphi
_{\ell }\right] ,  \nonumber
\end{eqnarray}
where the limits of integration are given by
\begin{equation}
\cos {\tilde\theta} _{\ell Q}=\max \left[ -1,\frac{1}{\alpha }\left( 1-\frac{%
m_{Q}^{2}(1-\delta ^{2})}{2E_{\ell }E_{Q}}\right) \right]  \label{17}
\end{equation}
and
\begin{equation}
{\tilde E}_{Q}=\frac{m_{Q}}{\sqrt{1-\max^{2}[\tilde \alpha,0]}},\qquad \tilde \alpha =%
\frac{4E_{\ell }^{2}-m_{Q}^{2}(1-\delta ^{2})}{4E_{\ell
}^{2}+m_{Q}^{2}(1-\delta ^{2})}.  \label{18}
\end{equation}

Note that Eqs.(\ref{16})-(\ref{18}) are applicable not only at the
Born level, but also in all those cases when radiative corrections
have a factorizable form (\ref{8}). At the tree level,
Eq.(\ref{16}) can be simplified due to $\delta $-function in the
Born cross sections (\ref{9}) and (\ref{10}).

\subsection{Hadron Level Results at LO}

At the hadron level, the overall one-particle inclusive kinematical invariants
of the reaction (\ref{4}) are defined as
\begin{eqnarray}
U &=&\left( k_{\gamma }-p_{\ell }\right) ^{2}=-E_{\ell }\sqrt{zS}(1-\cos
\theta _{\ell }),\qquad \ \ \ S=\left( k_{\gamma }+p_{N}\right) ^{2},
\nonumber \\
T &=&\left( p_{N}-p_{\ell }\right) ^{2}=-E_{\ell }\sqrt{S/z}(1+\cos \theta
_{\ell }),\qquad k_{g}=zp_{N}.  \label{19}
\end{eqnarray}
The hadron level cross section has the form of a convolution:
\begin{equation}
E_{\ell }\frac{\text{d}^{3}\sigma _{\ell }}{\text{d}^{3}p_{\ell }}\left(
S,T,U,\varphi _{\ell }\right) =\int\limits_{z_{m}}^{1}\text{d}zg(z,\mu _{F})%
\frac{E_{\ell }\text{d}^{3}\hat{\sigma}_{\ell }}{\text{d}^{3}p_{\ell }}%
\left( zS,zT,U,\varphi _{\ell }\right) ,  \label{20}
\end{equation}
in which $g(z,\mu _{F})$ is the gluon distribution function and $\mu _{F}$ is the
factorization scale. The lower limit of integration in Eq. (\ref{20}) can be
found from the kinematical restriction $x\leq 1-\delta ^{2}$. The result is:
\begin{equation}
z_{m}=\frac{4m_{Q}^{2}/S}{1-\max^{2}[\beta _{-},0]},  \label{21}
\end{equation}
where
\begin{equation}
\beta _{-}=\frac{m_{Q}^{2}(1-\delta ^{2})}{U}\left[ 1-\sqrt{\left( 1+\frac{U%
}{m_{Q}^{2}(1-\delta ^{2})}\right) ^{2}+\frac{4UT}{m_{Q}^{2}S(1-\delta
^{2})^{2}}}\right] .  \label{22}
\end{equation}

Experimentally, to suppress the background contributions originating from
semileptonic decays of light hadrons and Bethe-Heitler process, a cut of the
lepton energy in the reaction (\ref{4}) is usually used \cite{11}. For this reason,
apart from the usual differential distribution,
\begin{equation}
\frac{\text{d}^{2}\sigma _{\ell }}{\text{d}p_{\ell T}\text{d}\varphi _{\ell }%
}(p_{\ell T},\varphi _{\ell })=\int\limits_{p_{\ell T}}^{E_{\ell ,\max }^{*}}%
\frac{E_{\ell }\text{d}^{3}\sigma _{\ell }}{\text{d}^{3}p_{\ell }}\left(
S,T,U,\varphi _{\ell }\right) \frac{\lambda _{-}\text{d}E_{\ell }^{*}}{2%
\sqrt{E_{\ell }^{*2}-p_{\ell T}^{2}}},  \label{23}
\end{equation}
we consider also the quantity
\begin{equation}
\frac{\text{d}^{2}\sigma _{\ell }}{\text{d}p_{\ell T}\text{d}\varphi _{\ell }%
}(E_{\ell ,\rm{cut}}^{*},p_{\ell T},\varphi _{\ell
})=\int\limits_{E_{\ell ,\rm{cut}}^{*}}^{E_{\ell ,\max }^{*}}\frac{%
E_{\ell }\text{d}^{3}\sigma _{\ell }}{\text{d}^{3}p_{\ell }}\left(
S,T,U,\varphi _{\ell }\right) \frac{\lambda _{-}\text{d}E_{\ell }^{*}}{2\sqrt{%
E_{\ell }^{*2}-p_{\ell T}^{2}}}.  \label{24}
\end{equation}
In Eqs. (\ref{23}) and (\ref{24}), $E_{\ell }^{*}$ is the lepton energy in the lab
(nucleon rest) frame,
\begin{equation}
E_{\ell }^{*}=\frac{m_{N}^{2}-T}{2m_{N}},  \label{25}
\end{equation}
with the maximum value
\begin{equation}
E_{\ell ,\max }^{*}=\frac{\sqrt{S}}{8m_{N}}\left( \lambda _{+}\sqrt{S}%
(1+\beta )(1-\delta ^{2})+\lambda _{-}\sqrt{S(1+\beta )^{2}(1-\delta
^{2})^{2}-16p_{\ell T}^{2}}\right)  \label{26}
\end{equation}
and an experimental cutoff $E_{\ell ,\rm{cut}}^{*}$; $\lambda _{\pm
}=1\pm m_{N}^{2}/S$ and $\beta =\sqrt{1-4m_{Q}^{2}/S}$.

Let us discuss the hadron level pQCD predictions for the asymmetry in
azimuthal distribution of the decay lepton. In this paper, we will discuss
only the charm photoproduction at the SLAC\ energy $E_{\gamma }\approx$ 35 GeV, with $%
E_{\gamma }=(S-m_{N}^{2})/2m_{N}$. Unless otherwise stated, the CTEQ5M \cite{12}
parametrization of the gluon distribution function is used. The default
value of the charm quark mass is $m_{c}=$ 1.5 GeV.

\begin{figure}
\mbox{\epsfig{file=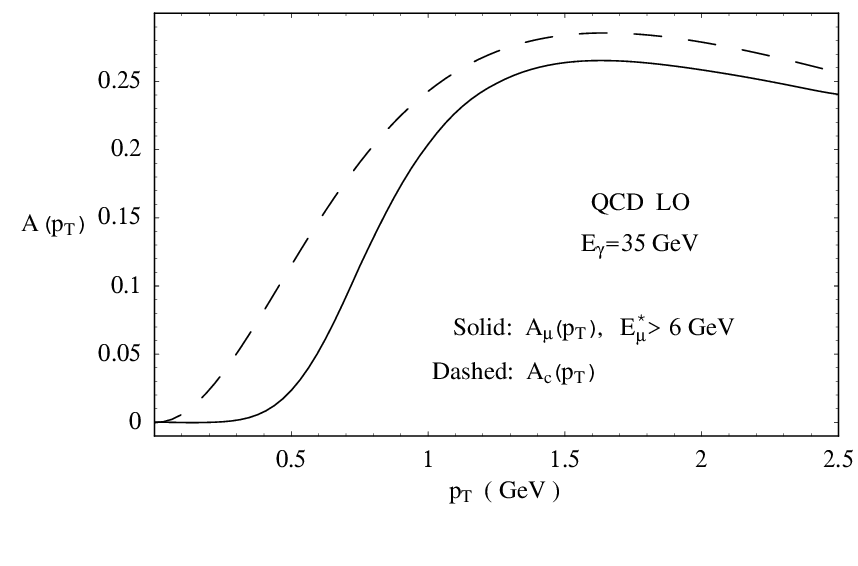}}
\caption{\label{Fig.1}
Comparison of the QCD LO pedictions for $A_{\mu }(p_{T})$ and
 $A_{c}(p_{T})$.}
\end{figure}

Our calculations of the quantities $A_{\mu }(p_{T})$ and
$A_{c}(p_{T})$ are given in Fig.\ref{Fig.1} by solid and dashed
lines, respectively. One can see that the asymmetry transferred
from the decaying $c$-quark to the decay muon is large in the SLAC
kinematics; the ratio $A_{\mu }(p_{T})/A_{c}(p_{T})$ is about
$90\%$ for $p_{T}>$1 GeV. Note that $p_{T}\equiv p_{QT}$ when we
consider the heavy quark production and $p_{T}\equiv p_{\ell T}$
when the quantity $A_{\mu }(p_{\ell T})$ is discussed.

\begin{figure}
\mbox{\epsfig{file=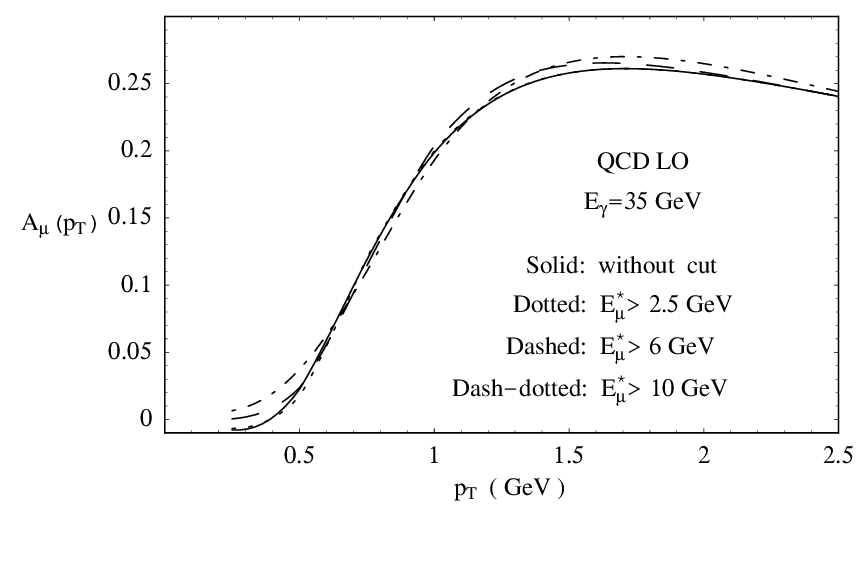}}
\caption{\label{Fig.2}
Dependence of the leptonic SSA, $A_{\mu }(p_{T})$,
on the cutoff parameter $E_{\mu ,\rm{cut}}^{*}$.}
\end{figure}

\begin{figure}
\mbox{\epsfig{file=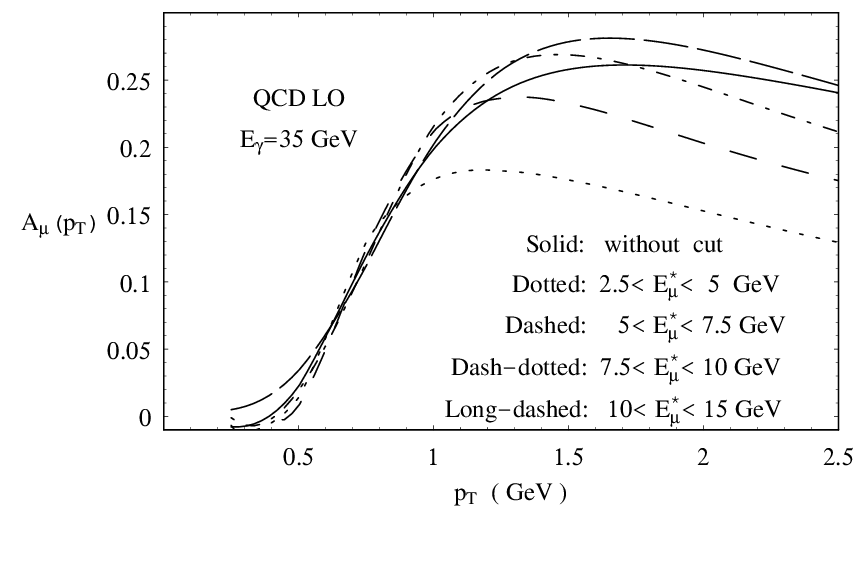}} \caption{\label{Fig.8} SSA,
$A_{\mu }(p_{T})$, in different intervals of the lepton energy
$E_{\mu}^{*}$.}
\end{figure}

In Fig. \ref{Fig.2} we show the LO predictions for $A_{\mu
}(p_{T})$ at different values of the cutoff parameter $E_{\ell
,\rm{cut}}^{*}$. One can see that, contrary to the production
cross sections, the SSA depends weakly on $E_{\ell ,\rm{cut}}^{*}$ 
over practically the whole region of $p_{T}$. This property of
$A_{\mu }(p_{T})$ will be especially useful in interpretation of
data. In Fig. \ref{Fig.8} we show the SSA, $A_{\mu }(p_{T})$, in
different intervals of the muon energy $E_{\ell}^{*}$.

The most interesting property of the azimuthal asymmetry, closely related
to fast perturbative convergence, is its parametric stability\footnote{%
Of course, parametric stability of the fixed-order results does
not imply a fast convergence of the corresponding series. However,
a fast convergent series must be parametricaly stable. In
particular, it must be $\mu _{R}$- and $\mu _{F}$-independent.}.
As was shown in \cite{9}, the LO predictions for $
A_{Q}(p_{T})$ are insensitive (to within few percent) to standard
theoretical uncertainties in the QCD input parameters: $\mu _{R}$,
$\mu _{F}$, $\Lambda _{QCD}$ and in the gluon distribution
function. We have verified that the same situation takes place in
the case of decay leptons too. In particular, all the CTEQ5
versions of the gluon density, as well as the MRST \cite{13}
parametrizations, lead to predictions for $A_{\mu }(p_{T})$ that 
coincide with each other with an accuracy of better than
1-2$\%$.

\begin{figure}
\begin{center}
\mbox{\epsfig{file=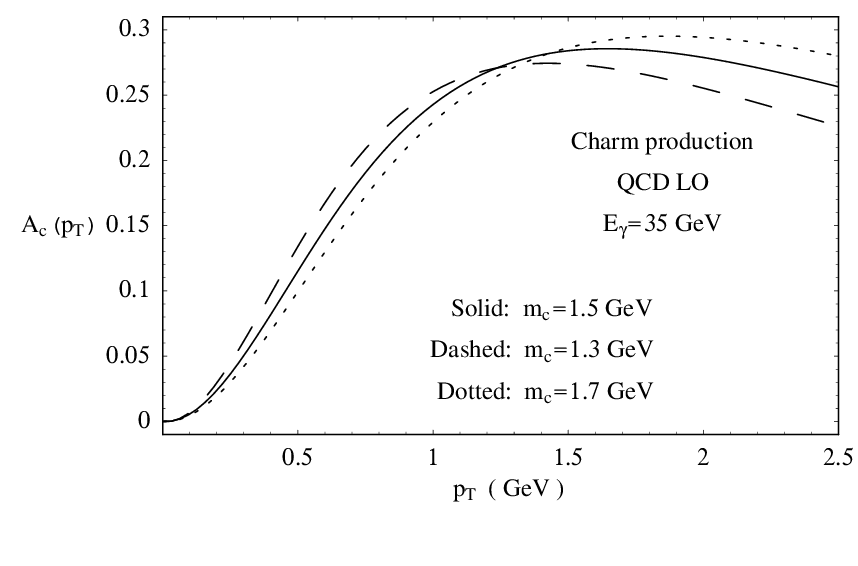}}
\end{center}
\caption{\label{Fig.3}
Dependence of the SSA in charm production, $A_{c }(p_{T})$,
on the $c$-quark mass, $m_{c}$.}
\end{figure}

\begin{figure}
\mbox{\epsfig{file=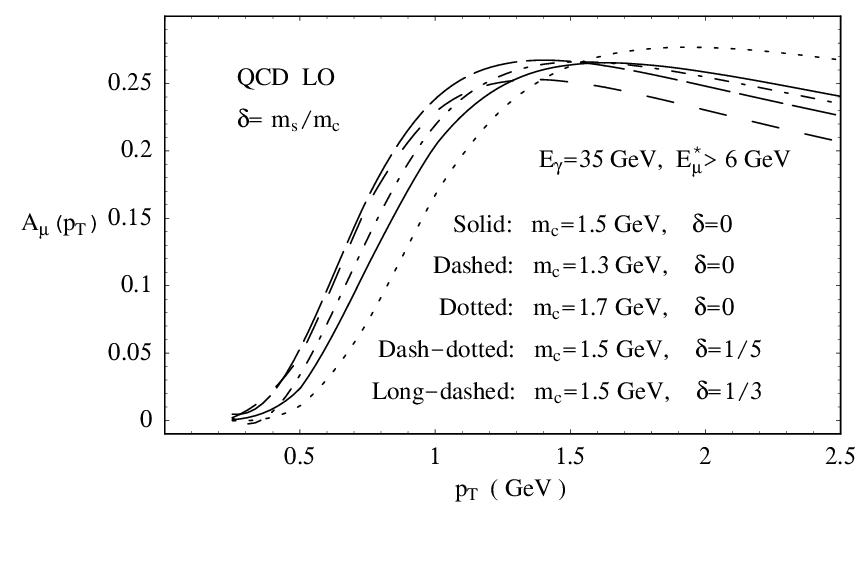}}
\caption{\label{Fig.4}
Dependence of the leptonic SSA, $A_{\mu }(p_{T})$,
on the charm and strange quark mass,  $m_{c}$ and  $m_{s}$.}
\end{figure}

The main source of uncertainties in the LO predictions for both
$A_{c}(p_{T}) $ and $A_{\mu }(p_{T})$ is the charm quark mass. The
dependence of $p_{T}$-spectra on the variation of $m_{c}$ is shown
in Fig. \ref{Fig.3} and Fig. \ref{Fig.4} for $A_{c}(p_{T})$ and
$A_{\mu }(p_{T})$, respectively. One can see that changes of the
charm quark mass in the interval $1.3<m_{c}<1.7$ GeV affect the
quantity $A_{c}(p_{T})$ by less than 20\% at $p_{T}<2.5$ GeV.
Analogous changes of $m_{c}$ lead to $20\%$ variations of $A_{\mu
}(p_{T})$ at $ 1<p_{T}<2.5$ GeV.

We have also analyzed the dependence of the SSA in the lepton
distribution on the unobserved strange quark mass, $m_{s}$. 
Fig. \ref{Fig.4} shows that the LO predictions for $A_{\mu
}(p_{T})$ are practically independent of $\delta=m_s/m_c$ at 
$p_{T}>$1 GeV.

\subsection{Radiative Corrections}

In Ref. \cite{10}, radiative corrections to the SSA in heavy quark
production by linearly polarized photons have been investigated in
the soft-gluon approximation. The calculations show that the
azimuthal asymmetry in both charm and bottom production is
practically insensitive to the soft-gluon corrections at energies
of the fixed target experiments. This implies that large
soft-gluon contributions to the spin-dependent and unpolarized
cross sections cancel each other in Eq. (\ref{3}) with a good
accuracy. One can assume that the same situation takes place 
for the azimuthal asymmetry in the decay lepton distribution.

Our calculations of the factorizable NLO corrections to $A_{\mu
}(p_{T})$ show that this is really the case. We have computed both
spin-dependent and unpolarized differential distributions
(\ref{9}) of the heavy-quark photoproduction at NLO to the
next-to-leading logarithmic accuracy. The NLO corrections to the
width of the heavy quark semileptonic decays are known exactly
\cite{A1,A2}. We have found that radiative corrections to the
leptonic SSA, $A_{\mu }(p_{T})$, in the reaction (\ref{4}) are of
the order of $1\%$ in the SLAC kinematics.

Two main reasons are responsible for perturbative stability of the
quantity $A_{\mu }(p_{T})$. First, radiative corrections to the
SSA in heavy quark production are small \cite{10}. Second, the
ratio $I_{\rm{sl}}^{\rm{NLO}}(x)/ I_{\rm{sl}}^{\rm{Born}}(x)$ is a
constant practically at all $x$, except for a narrow endpoint
region $x\approx 1$ \cite{A2}. (Note that
$I_{\rm{sl}}^{\rm{Born}}(x)$ is the LO invariant width of the
semileptonic decay $c\rightarrow \ell ^{+}\nu _{\ell }X_{q}$ given
by (\ref{13}) while $I_{\rm{sl}}^{\rm{NLO}}(x)$ is the
corresponding NLO one.) The details of our NLO analysis are too
long to be presented here and will be given in a separate
publication \cite{14}.

\section{Nonperturbative Contributions}

Let us discuss how the pQCD predictions for single spin asymmetry
are affected by nonperturbative contributions due to the intrinsic
transverse motion of the gluon and the hadronization of the
produced heavy quark. Because of the relatively low $c$-quark mass, 
these contributions are especially important in the description of the
cross section for charmed particle production \cite{1}. At the
same time, our analysis shows that nonperturbative corrections to
the single spin asymmetry are not large.

\subsection{Fragmentation}

Hadronization effects in heavy flavor production are usually modeled with
the help of the Peterson fragmentation function \cite{15},
\begin{equation}
D(y)=\frac{a_{\varepsilon }}{y\left[ 1-1/y-\varepsilon /(1-y)\right] ^{2}},
\label{28}
\end{equation}
where $a_{\varepsilon }$ is a normalization factor and
$\varepsilon _{D}=0.06$ in the case of a $D$-meson production. The
double differential distribution of a heavy meson has the form
\begin{equation}
\frac{\text{d}^{2}\sigma _{D}}{\text{d}p_{DT}\text{d}\varphi _{Q}}({\vec{p}}%
_{DT})=\int\limits_{y_{m}}^{1}\frac{\text{d}y}{y}D(y)\frac{\text{d}^{2}\sigma _{Q}}{%
\text{d}p_{QT}\text{d}\varphi _{Q}}\left( {\vec{p}}_{DT}/y\right) ,
\label{29}
\end{equation}
where $\frac{\text{d}^{2}\sigma
_{Q}}{\text{d}p_{QT}\text{d}\varphi _{Q}}$ is the inclusive
spectrum of the reaction (\ref{1}) and $y_{m}=
\frac{2p_{DT}}{\beta \sqrt{S}}$.

\begin{figure}
\begin{center}
\mbox{\epsfig{file=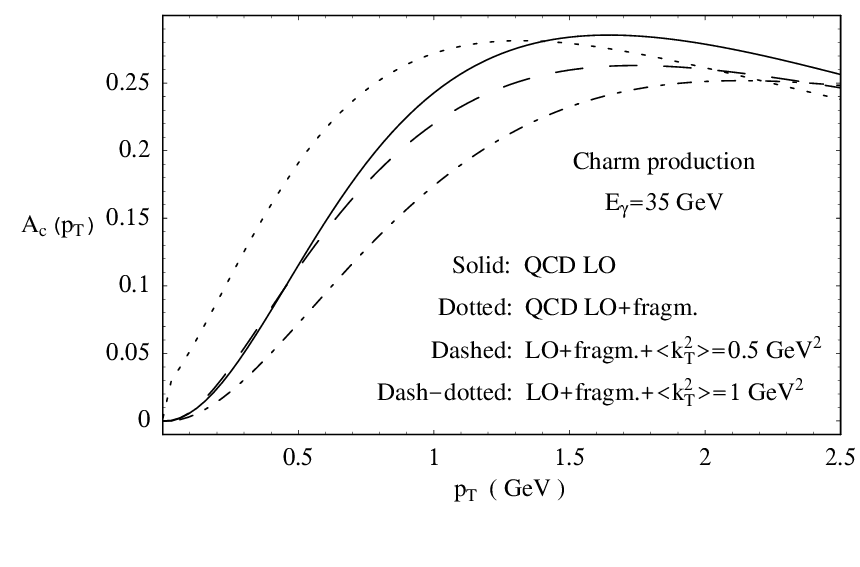}}
\end{center}
\caption{\label{Fig.5}
SSA in a $D$-meson production; the QCD LO predictions with
and without the inclusion of the $k_{T}$ smearing and Peterson
fragmentation effects.}
\end{figure}

Our calculations of the asymmetry in a $D$-meson production at LO
with and without the Peterson fragmentation effect are presented
in Fig. \ref{Fig.5} by dotted and solid curves, respectively. For 
$ p_{DT}\ge 1$ GeV the fragmentation corrections to
$A_{c}(p_{T})$ are less than $10\%$.

\begin{figure}
\begin{center}
\mbox{\epsfig{file=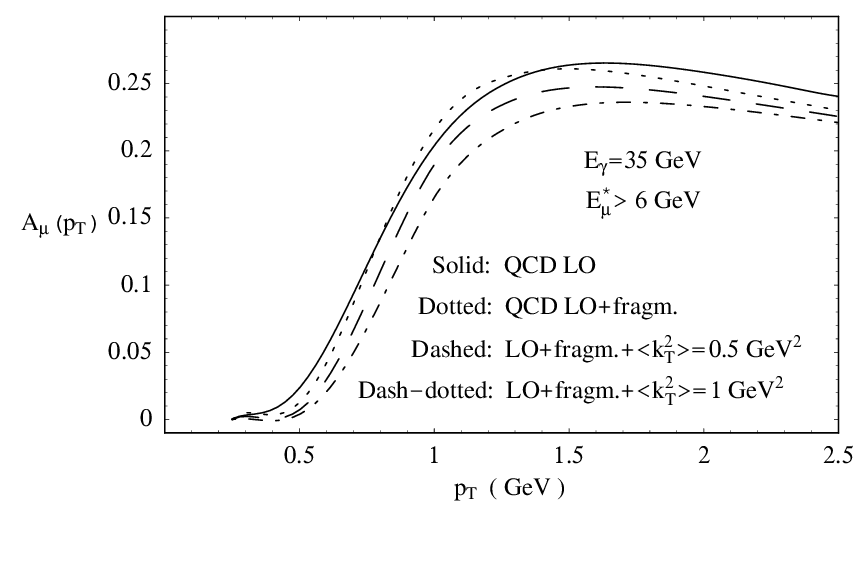}}
\end{center}
\caption{\label{Fig.6} SSA, $A_{\mu }(p_{T})$, in the decay lepton
distribution; the QCD LO predictions with and without the
inclusion of the $k_{T}$ smearing and Peterson fragmentation
effects.}
\end{figure}

Analogous corrections to the asymmetry in the decay lepton
azimuthal distribution, $A_{\mu }(p_{T}),$ are given in
Fig. \ref{Fig.6}. One can see that the effect of the fragmentation
function (\ref{28}) is practically negligible in the whole region
of $p_{\ell T}$.

\subsection{k$_{T}$ Smearing}

To introduce $k_{T}$ degrees of freedom, $\vec{k}_{g}\simeq z\vec{k}_{N}+%
\vec{k}_{T}$, one extends the integral over the parton distribution function
in Eq. (\ref{20}) to $k_{T}$-space,
\begin{equation}
\text{d}zg(z,\mu _{F})\rightarrow \text{d}z\text{d}^{2}k_{T}f\left( \vec{k}%
_{T}\right) g(z,\mu _{F}).  \label{30}
\end{equation}
The transverse momentum distribution, $f\left( \vec{k}_{T}\right) $, is
usually taken to be a Gaussian:
\begin{equation}
f\left( \vec{k}_{T}\right) =\frac{{\rm{e}}^{-k_{T}^{2}/\langle
k_{T}^{2}\rangle }}{\pi \langle k_{T}^{2}\rangle }.  \label{31}
\end{equation}
In practice, an analytic treatment of $k_{T} $ effects is usually
used. According to \cite{16}, the $k_{T}$-smeared differential
cross section of the process (\ref{1}) is a 2-dimensional
convolution:
\begin{equation}
\frac{\text{d}^{2}\sigma _{Q}^{\rm{kick}}}{\text{d}p_{QT}\text{d}\varphi
_{Q}}\left( {\vec{p}}_{QT}\right) =\int \text{d}^{2}k_{T}\frac{{\rm{e}}%
^{-k_{T}^{2}/\langle k_{T}^{2}\rangle }}{\pi \langle k_{T}^{2}\rangle }\frac{%
\text{d}^{2}\sigma _{Q}}{\text{d}p_{QT}\text{d}\varphi _{Q}}\left( {\vec{p}}%
_{QT}-\frac{1}{2}\vec{k}_{T}\right) .  \label{32}
\end{equation}
The factor $\frac{1}{2}$ in front of $\vec{k}_{T}$ in the r.h.s. of Eq. (\ref{32}%
) reflects the fact that the heavy quark carries away about one half of the
initial energy in the reaction (\ref{1}).

Values of the $k_{T}$-kick corrections to the asymmetry in the
charm production, $A_{c}(p_{T})$, are shown in Fig. \ref{Fig.5} by
dashed ($\langle k_{T}^{2}\rangle =0.5$ GeV$^{2}$) and dash-dotted
($\langle k_{T}^{2}\rangle =1$ GeV$^{2}$) curves. One can see that
$k_{T}$-smearing is important only in the region of relatively low
$p_{QT}\leq m_{c}$. Note also that the fragmentation and
$k_{T}$-kick effects practically cancel each other in the case of
$\langle k_{T}^{2}\rangle =0.5$ GeV$^{2}$.

Corresponding calculations for the case of the lepton asymmetry
are presented in Fig. \ref{Fig.6}. It is seen that $A_{\mu
}(p_{T})$ is affected by $k_{T}$-corrections systematically
less than $A_{c}(p_{T})$.

\subsection{Fermi Motion}

The third type of nonperturbative corrections we considered can be
associated with the motion of the heavy quark inside the produced (and
decaying) hadron; they are commonly referred to as Fermi motion. These
effects are included in the heavy-quark expansion by resumming an infinite
set of leading-twist corrections into a shape function $F\left( k_{+}\right)
$, where $k_{+}=k_{0}+k_{3}$ is the positive light cone component of the
heavy quark residual momentum inside the heavy meson \cite{17,18}. The
physical decay distributions, d$\Gamma _{\rm{sl}}^{\rm{(H)}}$, are obtained from
a convolution of parton model spectra, d$\Gamma _{\rm{sl}}^{\rm{(Q)}}$, with
this function. This convolution is such that, in the perturbative formula
for the decay distribution, the heavy quark mass is replaced by the 
momentum-dependent mass $m^{*}=m_{Q}+$ $k_{+}$ \cite{19}:
\begin{equation}
E_{\ell }\frac{\text{d}^{3}\Gamma _{\rm{sl}}^{\rm{(H)}}}{\text{d}^{3}p_{\ell }%
}(M_{H},x_{H})=\int\limits_{x_{H}M_{H}}^{M_{H}}\frac{E_{\ell }\text{d}%
^{3}\Gamma _{\rm{sl}}^{\rm{(Q)}}}{\text{d}^{3}p_{\ell }}\left( m^{*},x_{H}%
\frac{M_{H}}{m^{*}}\right) F\left( m^{*}\right) \text{d}m^{*}.  \label{33}
\end{equation}
The parton-level width, d$^{3}\Gamma _{\rm{sl}}^{\rm{(Q)}}(m_{Q},x)$, 
in the r.h.s of Eq.(\ref{33}) is defined by Eq. (\ref{13}), $M_{H}$ is the
heavy meson mass and the hadron level variable $x_{H}$ is 
related to $x=2(p_{\ell }\cdot p_{Q})/m_{Q}^{2}$ by 
$x_{H}=xm_{Q}/M_{H}$.

Some of the properties of $F\left( k_{+}\right) $ are known. The first three
moments satisfy $\Lambda _{0}=1$, $\Lambda _{1}=0$ and $\Lambda _{2}=\frac{1%
}{3}\mu _{\pi }^{2}$, where $\mu _{\pi }^{2}$ is the average momentum
squared of the heavy quark inside the heavy meson and the moments are
defined by $\Lambda _{n}=\langle k_{+}^{n}\rangle =\int_{-m_{Q}}^{\overline{%
\Lambda }}$d$k_{+}k_{+}^{n}F\left( k_{+}\right) $ with $\overline{\Lambda }%
=M_{H}-m_{Q}$. In our calculations we use a simple two-parametric
model \cite{19}:
\begin{equation}
F\left( k_{+}\right) =N{\rm{e}}^{(1+a)\eta }(1-\eta )^{a};\qquad
\qquad \qquad \eta =\frac{k_{+}}{\overline{\Lambda }},  \label{34}
\end{equation}
where the condition $\Lambda _{0}=1$ fixes the normalization
factor $N$ and the parameter $a$ is related to the second moment
as $\Lambda _{2}=\frac{1}{3}\mu _{\pi }^{2}=\overline{\Lambda
}^{2}/(1+a)$.

\begin{figure}
\begin{center}
\mbox{\epsfig{file=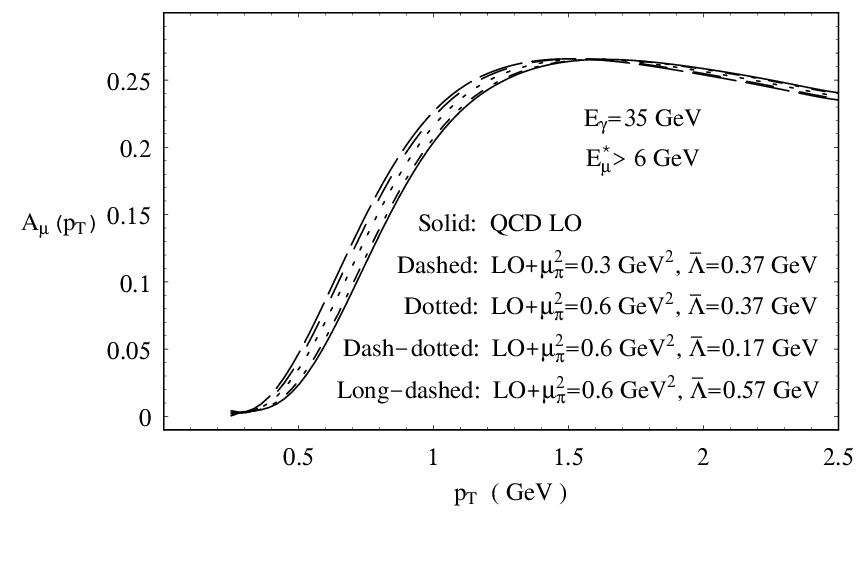}}
\end{center}
\caption{\label{Fig.7} Fermi motion corrections to the leptonic
SSA, $A_{\mu }(p_{T})$, at different values of $\mu _{\pi }^{2}$
and $\overline{\Lambda }$.}
\end{figure}

Our predictions for the asymmetry $A_{\mu }(p_{T})$ at different
values of $\mu _{\pi }^{2}$ and $\overline{\Lambda }$ are give in
Fig. \ref{Fig.7}. One can see that the bound state effects due to
the Fermi motion of the heavy quark inside the heavy hadron are
small in the whole region of $p_{\ell T}$. Taking into account
that $A_{\mu }(p_{T})$ is also practically independent of $m_{s}$
at $p_{\ell T}\ge 1$ GeV, we conclude that our predictions for the
SSA in secondary lepton distribution are insensitive to the
theoretical uncertainties in description of the charm semileptonic
decays \cite{21}.

\section{Experimental Considerations}
The conditions needed to measure the SSA in open charm
photoproduction are:
\begin{itemize}
\item a high intensity photon beam with energy above 30
GeV,
\item a high degree of linear polarization, and
\item a large acceptance spectrometer.
\end{itemize}
These conditions are similar to those of the approved E161 experiment
\cite{11} at SLAC, in which circularly polarized photons impinging
on longitudinally polarized nucleons will be used to study $\Delta
G(x)$ via open charm production. To measure the SSA under
consideration, target polarization is not needed, so the
microwaves that polarize the planned LiD target can be turned off.
Linearly polarized photons can be obtained with an appropriate choice
of the diamond used to produce coherent bremsstrahlung. Two
orientations of linear polarization are obtained by rotating
the diamond. The SSA can be measured using high $p_T$ muons in the
E161 spectrometer, with cuts that minimize backgrounds.

\subsection{Photon Beam}
\begin{figure}
\begin{center}
\mbox{\epsfig{file=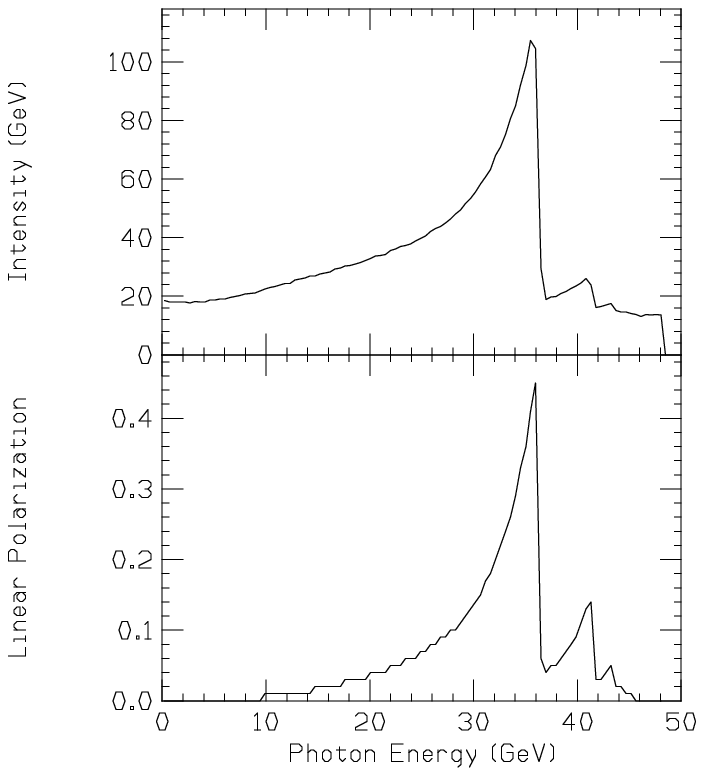}}
\end{center}
\caption{\label{Fig.9} The upper panel shows the predicted
relative intensity of photons (intensity is energy-weighted flux)
for coherent bremsstrahlung with a primary coherent peak at
$E_{\gamma}=35$ GeV. The lower panel shows the calculated linear
polarization under conditions typical of SLAC E161.}
\end{figure}

The most important difference from E161 is the change from
circular to linear polarization. The E161 experiment plans to use
coherent bremsstrahlung from a diamond target to produce polarized
photons. Longitudinally polarized electrons are needed for
circular polarization, which is largest for large values of
$E_{\gamma}/E_e$, where $E_e$ is the incident electron beam
energy, and $E_{\gamma}$ is the resulting photon beam energy. To
obtain linear polarization, the electrons can be unpolarized, and
the extent of linear polarization is largest for low values of
$E_{\gamma}/E_e$. For a fixed photon energy, it is therefore
essential to use the highest possible value of $E_e$, which is
about 50 GeV at SLAC. Traditionally, thin diamonds and a high
degree of photon collimation are used to enhance the ratio of
coherent to incoherent photons, and hence the effective
polarization. To obtain a sufficient flux of photons at SLAC, it
is necessary to use a relatively thick diamond (we choose 1.5 mm),
and a modest collimation angle of 35 $\mu$r, which is several times the
characteristic bremsstrahlung angle $m_{e}/E_{e} \approx 10$
$\mu$r. The photon intensity and the degree of linear polarization for
a diamond orientation that produces a primary peak from the $(02
\bar{2})$ inverse lattice point at 35 GeV are shown in
Fig. \ref{Fig.9} for the emittance and maximum intensity conditions
of the SLAC electron beam. Calculations were done using a Monte
Carlo simulation based on the formulas of Ref.~\cite{palazzi}. The
flux in the region 30 to 35 GeV is about $10^{10}$ photons/sec.
The linear polarization peaks at about 0.4, with an average value
of about 0.2 in the 30 to 35 GeV region. Contributions from the
higher energy peaks (from the $(04 \bar{4})$, $(06 \bar{6})$, etc.
inverse lattice points) will be small because both the flux and
polarization are considerably reduced compared to the primary
coherent peak.

\subsection{Open Charm Detection and Backgrounds}
Because of the poor duty factor at SLAC, simulations made for E161
have shown that the best signal-to-background ratio is obtained by
tagging open charm with one or more high transverse momentum
muons. In the case of the SSA, the asymmetry rises rapidly with
$p_T$, then becomes roughly constant for $p_T>1$ GeV. All
background processes except $J/\psi$ decay decrease more rapidly
with $p_T$ than open charm (as discussed in more detail below),
therefore measuring at the highest possible $p_T$ will optimize
signals over backgrounds. However, the open charm cross section
decreases with $p_T$, which reduces statistical accuracy. The best
compromise is for $p_T$ values centered at 1 GeV.

\begin{figure}
\begin{center}
\mbox{\epsfig{file=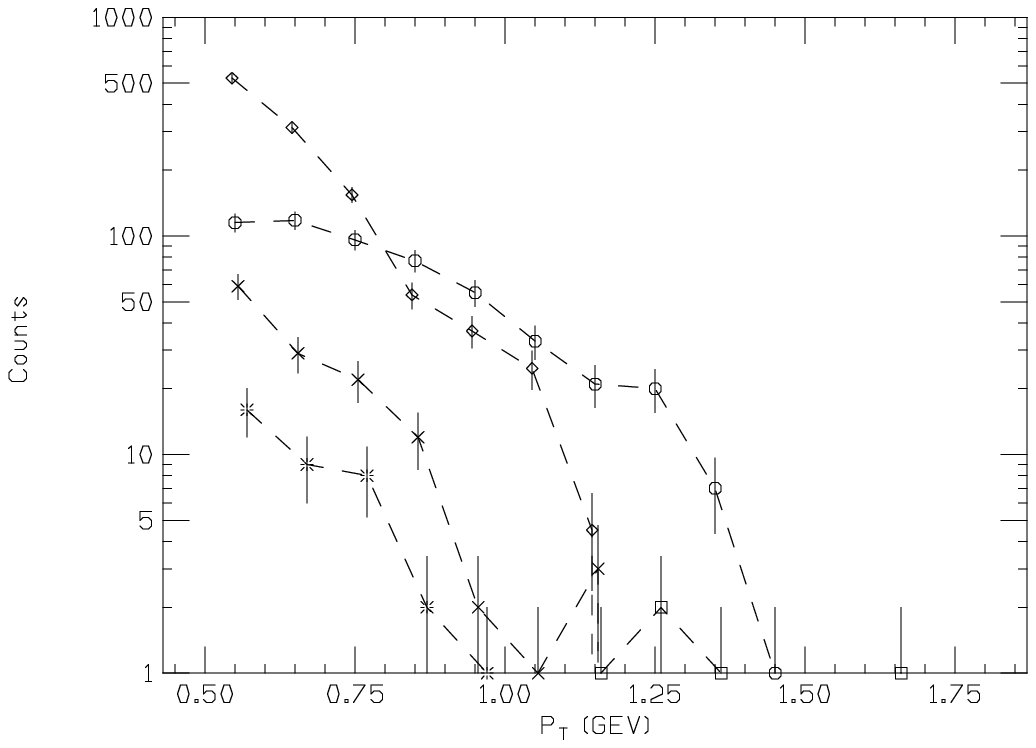}}
\end{center}
\caption{\label{Fig.10} Relative number of counts from open charm
(circles), pion and kaon decays (diamonds), nuclear pair
production (crosses), light vector meson decays (stars), and
$J/\psi$ decays (boxes), for $7.5<E_{\mu}^{*}<10$ GeV as a
function of $p_T$ for $30<E_{\gamma}<35$ GeV. The count rates
correspond to about one minute under E161 conditions.}
\end{figure}

The most significant background process generating single muon with 
high $p_T$ is the decay of pions and kaons. This was simulated using
PYTHIA \cite{pythia} for typical photoproduction events 
that were allowed to decay in a block of copper placed close 
to the target. Studies using GEANT indicate that an effective path
length before pions and kaons are absorbed is 50 cm of copper, so
this was used as the effective decay length. The rates from $\pi,K$
decays are shown as the diamonds in Fig. \ref{Fig.10} for any sign
muon with $7.5<E_{\mu}^{*}<10$ GeV. They are compared with the 
rate of muons predicted from open charm decay as calculated using the
photon-gluon fusion process in PYTHIA, scaled by a factor of two
to match experimental data. 
The $\pi,K$ decay background is lower for negative muons than for
positive muons by typically 50\%, so one sign will be easier to
measure than the other. 
The rate and SSA for $\pi,K$ decay muons can be measured by moving
the copper absorber a significant distance from the target, thus
doubling or tripling the decay rates. Measurements can also be
made for $5<E_{\mu}^{*}<7.5$ GeV, with somewhat larger relative
backgrounds from $\pi,K$ decays. In the interesting range
$0.8<p_T<1.2$ GeV, $\pi$ and $K$ decays become dominant below
$E_{\mu}^{*} \approx 4$ GeV, making experimental measurements
difficult in this energy region.

Another significant process is wide-angle muon pair production
from the target nuclei (Li and D). The rates calculated using
Ref. \cite{tsai} are shown as the crosses in Fig. \ref{Fig.10}. 
The rates are for events in which only one of the two muons from
the Bethe-Heitler pair is detected in the proposed E161
spectrometer. Events in which both muons would be detected can be
vetoed in the data analysis, reducing the Bethe-Heitler background
by about a factor of two. The remaining background is well below
the open charm rate. It can be calculated quite accurately, and
reliable corrections applied to the data.

Backgrounds from the decay of vector mesons are relatively
unimportant for $p_T \approx 1$ GeV, and can be measured using the
approximately 50\% of events in which both muons are detected in
the spectrometer.

The final background is from $J/\psi$ decay, for which the typical
$p_T$ peaks around 1.5 GeV. Fortunately, most of the time the
second muon is detected in the E161 spectrometer, so the
background only becomes large for $p_T>1.3$ GeV and
$E_{\mu}^{*}>10$ GeV, and can be measured using the large sample
of events in which both muons are detected.

In summary, in the optimal case of negative muons with 
$7.5<E_{\mu}^{*}<10$ GeV and $p_T>0.9$ GeV,  the
signal-to-background ratio is estimated to be about 3:1, leading
to an effective dilution factor of about 1.3, equivalent to a
factor of 1.6 more running time than is needed in the absence of
backgrounds. This estimate relies crucially on the open charm
cross section as a function of $p_T$, which is poorly known at
present (factor of two uncertainties). Thus, the relative
background situation could be better or worse, depending on what
open charm cross sections are measured in the experiment.

\subsection{Projected Errors}
The cross section for producing negative muons from open charm
decays with $7.5<E_{\mu}^{*}<10$ GeV and $p_T>0.9$ GeV is
approximately 0.2 to 0.5 nb at $E_{\gamma}=35$ GeV, as estimated
using the PYTHIA Monte Carlo. 
If we assume the E161 luminosity, i.e. a flux of  $10^{10}$
photons per second, a 1.6 gm/cm$^2$ target, and a spectrometer
acceptance of about 50\% for the events of interest, the open
charm rate is of order 100,000 events per day. For an average
linear polarization of 0.2, a statistical error of about 0.02 on
the SSA would be obtained in one day of running. This error bar is
sufficient to make a good test of the QCD predictions, and further
improvements to the statistical error would be soon be lost in the
estimated 10\% (relative) systematics error, dominated by the
uncertainty in the linear polarization. 
With about a factor of four larger running time, measurements with
an error of about 0.02 could be separately made of positive and
negative muons, and in several bins of $E_{\mu}^{*}$ spanning
$5<E_{\mu}^{*}<20$ GeV. This would be very useful in testing the
assumed dominance of the photon-gluon fusion mechanism. A
difference in positive and negative muon results might indicate
contributions from soft associated production processes, while
high-$x_F$ behavior of the SSA can be sensitive to diffractive
contributions \cite{9}. Approximately the same running time would
be needed for 25 GeV incident photons as for $E_{\gamma}=35$ GeV,
because the lower charm cross section and higher backgrounds would
be compensated for by higher photon linear polarization.

\section{Conclusion}
In this paper we analyze the possibility to measure the SSA in
open charm photoproduction in the E160/E161 experiments at SLAC
where a coherent bremsstrahlung beam of linearly polarized photons
with energies up to 40 GeV will be available soon. In these
experiments, charm production will be investigated with the
help of inclusive spectra of secondary muons. The SSA
transferred from the decaying $c$-quark to the decay muon is
predicted to be large for SLAC kinematics; the ratio $A_{\ell
}(p_{T})/A_{c}(p_{T})$ is about $90\%$ at $p_{T}>$1 GeV. Our
calculations show that the SSA in decay lepton distribution
preserves all remarkable properties of the SSA in heavy flavor
production: it is stable, both perturbatively and parametrically,
and practically insensitive to nonperturbative contributions due
to the gluon transverse motion in the target and heavy quark
fragmentation. We have also found that QCD predictions for
$A_{\ell}(p_{T})$ depend weekly on theoretical uncertainties in
the charm semileptonic decays. We conclude that measurements of
$A_{\ell}(p_{T})$ in the E160/E161 experiments would provide a
good test of pQCD applicability to open charm production.

Note also that because of the low $c$-quark mass, the power
corrections to the charm production can be essential. As was 
shown in \cite{9}, the pQCD and Regge approaches lead to strongly
different predictions for the single spin asymmetry in the region
of low $p_{T}$ and large Feynman $x_{F}$. Data on the $p_{T}$- and
$x_{F}$-distributions of the SSA in $D$-meson photoproduction 
could make it possible to discriminate between these mechanisms.

{\em Acknowledgements.} We would like to thank S.J. Brodsky, 
A. Capella, E. Chudakov, L. Dixon, A.B. Kaidalov,  A.E. Kuraev, 
M.E. Peskin and A.V. Radyushkin for useful discussions.
N.Ya.I. is grateful to Theory Groups of SLAC and JLab for
hospitality while this work has been completed. This work was
supported in part by the National Science Foundation.

\newpage

\vspace{5mm}

\newpage

\vspace{5mm}

\newpage

\vspace{5mm}

\newpage

\vspace{5mm}

\newpage

\newpage

\end{document}